\documentstyle[amsmath,amssymb,latexsym,epsfig]{KapEdbk}

\def \d{{\mathrm{d}}}

\def \pd{\partial}

\def \tl#1{\overset{\kern 1pt\circ}{#1}}
\def \TL#1{\overset{\kern -3pt \circ}{#1}}
\def \TLL#1{\overset{\kern -7pt \circ}{#1}}

\def \burger{{\bf b}}

\normallatexbib

\begin{document}
\articletitle
{Peach-Koehler forces within the theory of nonlocal elasticity}
\author{Markus Lazar}
\affil{Laboratoire de Mod{\'e}lisation en M{\'e}canique,\\
        Universit{\'e} Pierre et Marie Curie, \\
	4 Place Jussieu, Case 162,\\	
	F-75252 Paris Cedex 05, France}
\email{lazar@lmm.jussieu.fr}

\begin{abstract}
We consider dislocations in the framework of Eringen's nonlocal elasticity.
The fundamental field equations of nonlocal elasticity are presented. 
Using these equations, the nonlocal force stresses 
of a straight screw and a straight
edge dislocation are given.
By the help of these nonlocal stresses, we are able to calculate the
interaction forces between dislocations (Peach-Koehler forces). 
All classical singularities of the Peach-Koehler forces are eliminated. 
The extremum values of the forces are found near the dislocation line.\\
\end{abstract}


\section{Introduction}
Traditional methods of classical elasticity break down at small distances from 
crystal defects and lead to singularities. 
This is unfortunate since the defect core
is a very important region in the theory of defects.
Moreover, such singularities
are unphysical and an improved model of defects should eliminate them.
In addition, classical elasticity is a scale-free continuum theory 
in which no characteristic length appears.
Thus, classical elasticity cannot explain the phenomena near defects 
and at the atomic scale.

In recent decades, a theory of elastic continuum called nonlocal elasticity
has been developed. The concept of nonlocal elasticity was originally proposed 
by Kr{\"o}ner and Datta~\cite{KD66,Kroener67}, Edelen and Eringen~\cite{EE72,Eringen83,Eringen02}, 
Kunin~\cite{Kunin86} and some others. 
This theory considers the inner structures of materials and
takes into account long-range (nonlocal) interactions.

It is important to note that the nonlocal elasticity may be related to other
nonstandard continuum theories like gradient theory~\cite{GA99,Gutkin00} 
and gauge theory of defects~\cite{EL88,Lazar02b,Lazar03}. 
In all these approaches characteristic inner lengths (gradient coefficient or
nonlocality parameter), which describe size (or scale) effects, appear.
One remarkable feature of solutions in nonlocal elasticity, 
gradient theory and gauge theory 
is that the stress singularities which appear in classical elasticity 
are eliminated. 
These solutions depend on the characteristic inner length,
and they lead to finite stresses.
Therefore, they are applicable up to the atomic scale.

In this paper, we investigate the nonlocal force due to dislocations
(nonlocal Peach-Koehler force).
We consider parallel screw and edge dislocations.
The classical singularity of Peach-Koehler force is eliminated and 
a maximum/minimum is obtained. 

\section{Fundamental field equations}
The fundamental field equations for an isotropic, homogeneous, nonlocal 
and infinite extended medium with vanishing body force and static case
have been given by the nonlocal theory~\cite{EE72,Eringen83,Eringen02} 
\begin{align}
&\pd_j \sigma_{ij}=0,\nonumber\\
\label{stress-nl}
&\sigma_{ij}(r)=\int_V \alpha(r-r')\,\tl\sigma {}_{ij}(r')\, \d v(r'),
\nonumber\\
&\tl\sigma {}_{ij}=2\mu\left( \tl\epsilon {}_{ij}+\frac{\nu}{1-2\nu}\,\delta_{ij} \tl\epsilon {}_{kk}\right),
\end{align}
where $\mu$, $\nu$ are shear modulus and Poisson's ration, respectively.
In addition, $\tl\epsilon {}_{ij}$ is the classical strain tensor,
$\tl\sigma {}_{ij}$ and $\sigma_{ij}$ are the classical and nonlocal stress tensors,
respectively.
The $\alpha(r)$ is a nonlocal kernel.
The field equation in nonlocal elasticity of the stress in an isotropic medium 
is the following inhomogeneous Helmholtz equation
\begin{align}
\label{stress-fe}
\Big(1-\kappa^{-2}\Delta\Big)\sigma_{ij}=\tl\sigma {}_{ij},
\end{align}
where $\tl\sigma {}_{ij}$ is the stress tensor
obtained for the same traction boundary-value problem 
within the ``classical'' theory of dislocations.
The factor $\kappa^{-1}$ has the physical dimension of a length and 
it, therefore, defines an internal characteristic length.
If we consider the two-dimensional problem and
using Green's function of the two-dimensional Helmholtz equation~(\ref{stress-fe}), 
we may solve the field equation for every component of the stress 
field~(\ref{stress-fe}) by the help of the convolution integral
and the two-dimensional Green function
\begin{align}
\label{green}
\alpha(r-r')& =\frac{\kappa^2}{2\pi}\,K_0\big(\kappa \sqrt{(x-x')^2+(y-y')^2}\big).
\end{align}
Here $K_n$ is the modified Bessel function of the second kind and 
$n=0,1,\ldots$ denotes the order of this function.
Thus,
\begin{align}
\Big(1-\kappa^{-2}\Delta\Big)\alpha(r)=\delta(r),
\end{align}
where $\delta(r):=\delta(x)\delta(y)$ is the two-dimensional
Dirac delta function.
In this way, we deduce Eringen's so-called nonlocal constitutive relation
for a linear homogeneous, isotropic solid with Green's function~(\ref{green}) 
as the nonlocal kernel. This kernel~(\ref{green}) has its maximum at $r=r'$ and
describes the nonlocal interaction. 
Its two-dimensional volume-integral yields
\begin{align}
\int_V\alpha(r-r')\,\d v(r')=1,
\end{align}
which is the normalization condition of the nonlocal kernel.
In the classical limit ($\kappa^{-1}\rightarrow 0$), it becomes the Dirac delta function
\begin{align}
\lim_{\kappa^{-1}\to 0}\alpha(r-r')= \delta(r-r').
\end{align}
In this limit, Eq.~(\ref{stress-nl}) gives the classical expressions.
Note that Eringen~\cite{Eringen83,Eringen02} found the two-dimensional kernel~(\ref{green}) 
by giving the best match with the Born-K{\'a}rm{\'a}n model of the 
atomic lattice dynamics and the atomistic dispersion curves.
He used the choice $e_0=0.39$ for the length,  
$\kappa^{-1}=e_0\,a$,
where $a$ is an internal length (e.g. atomic lattice parameter)
and $e_0$ is a material constant.

\section{Nonlocal stress fields of dislocations}
Let us first review the nonlocal stress fields of screw and edge dislocations
in an infinitely extended body.
The dislocation lines are along the $z$-axis.
The nonlocal stress components of a straight screw dislocation with
the Burgers vector $\burger=(0,0,b_z)$ is given by~\cite{Eringen83,Eringen02,GA99,Lazar02b}
\begin{align}
\label{T-screw}
&\sigma_{xz}=-\frac{\mu b_z}{2\pi}\,\frac{y}{r^2}\Big\{1-\kappa r K_1(\kappa r)\Big\},\quad
\sigma_{yz}=\frac{\mu b_z}{2\pi}\,\frac{x}{r^2}\Big\{1-\kappa r K_1(\kappa r)\Big\},
\end{align}
where $r=\sqrt{x^2+y^2}$.
The nonlocal stress of a straight edge dislocation with the Burgers vector 
$\burger=(b_x,0,0)$ turns out to be~\cite{GA99,Lazar03}
\begin{align}
&\sigma_{xx}=-\frac{\mu b_x}{2\pi(1-\nu)}\, 
\frac{y}{r^4}\bigg\{\big(y^2+3x^2\big)
+\frac{4}{\kappa^2r^2}\big(y^2-3x^2\big)
-2 y^2\kappa r K_1(\kappa r)\nonumber\\
&\hspace{7.5cm}-2\big(y^2-3x^2\big) K_2(\kappa r)\bigg\},\nonumber\\
&\sigma_{yy}=-\frac{\mu b_x}{2\pi(1-\nu)}\, 
\frac{y}{r^4}\bigg\{\big(y^2-x^2\big)-\frac{4}{\kappa^2r^2}\big(y^2-3x^2\big)
-2 x^2\kappa r K_1(\kappa r)\nonumber\\
&\hspace{7.5cm}+2\big(y^2-3x^2\big) K_2(\kappa r)\bigg\},\nonumber\\
&\sigma_{xy}=\frac{\mu b_x}{2\pi(1-\nu)}\, 
\frac{x}{r^4}\bigg\{\big(x^2-y^2\big)-\frac{4}{\kappa^2r^2}\big(x^2-3y^2\big)
-2 y^2\kappa r K_1(\kappa r)\nonumber\\
&\hspace{7.5cm}+2\big(x^2-3y^2\big) K_2(\kappa r)\bigg\},\nonumber\\
\label{T_edge}
&\sigma_{zz}=-\frac{\mu b_x\nu }{\pi(1-\nu)}\, 
\frac{y}{r^2}\Big\{1-\kappa r K_1(\kappa r)\Big\}.
\end{align}
It is obvious that there is no singularity in~(\ref{T-screw}) and (\ref{T_edge}). 
For example, when $r$ tends to zero, the stresses $\sigma_{ij}\rightarrow 0$.
It also can be found that the far-field expression ($r>12\kappa^{-1}$) 
of Eqs.~(\ref{T-screw}) and (\ref{T_edge})  
return to the stresses in classical elasticity.
Of course, the stress fields~(\ref{T-screw}) and (\ref{T_edge}) fulfill 
Eq.~(\ref{stress-fe}) and correspond to the nonlocal kernel~(\ref{green}).

\section{Peach-Koehler forces due to dislocations}
The force between dislocations, according to Peach-Koehler
formula in nonlocal elasticity~\cite{KV79}, is given by
\begin{align}
\label{PKF}
F_k=\varepsilon_{ijk} \sigma_{in} b'_n \xi_j,
\end{align}
where $b'_n$ is the component of Burgers vector of the 2nd dislocation 
at the position $r$ and
$\xi_j$ is the direction of the dislocation.
Obviously, Eq.~(\ref{PKF}) is quite similar in the form as the 
classical expression of the Peach-Koehler force. 
Only the classical stress is replaced by the 
nonlocal one.
Eq.~(\ref{PKF}) is particularly important for the interaction between dislocations.

\subsection{Parallel screw dislocations}
\begin{figure}[t]\unitlength1cm
\vspace*{-1.0cm}
\centerline{
\begin{picture}(8,6)
\put(0.0,0.2){\epsfig{file=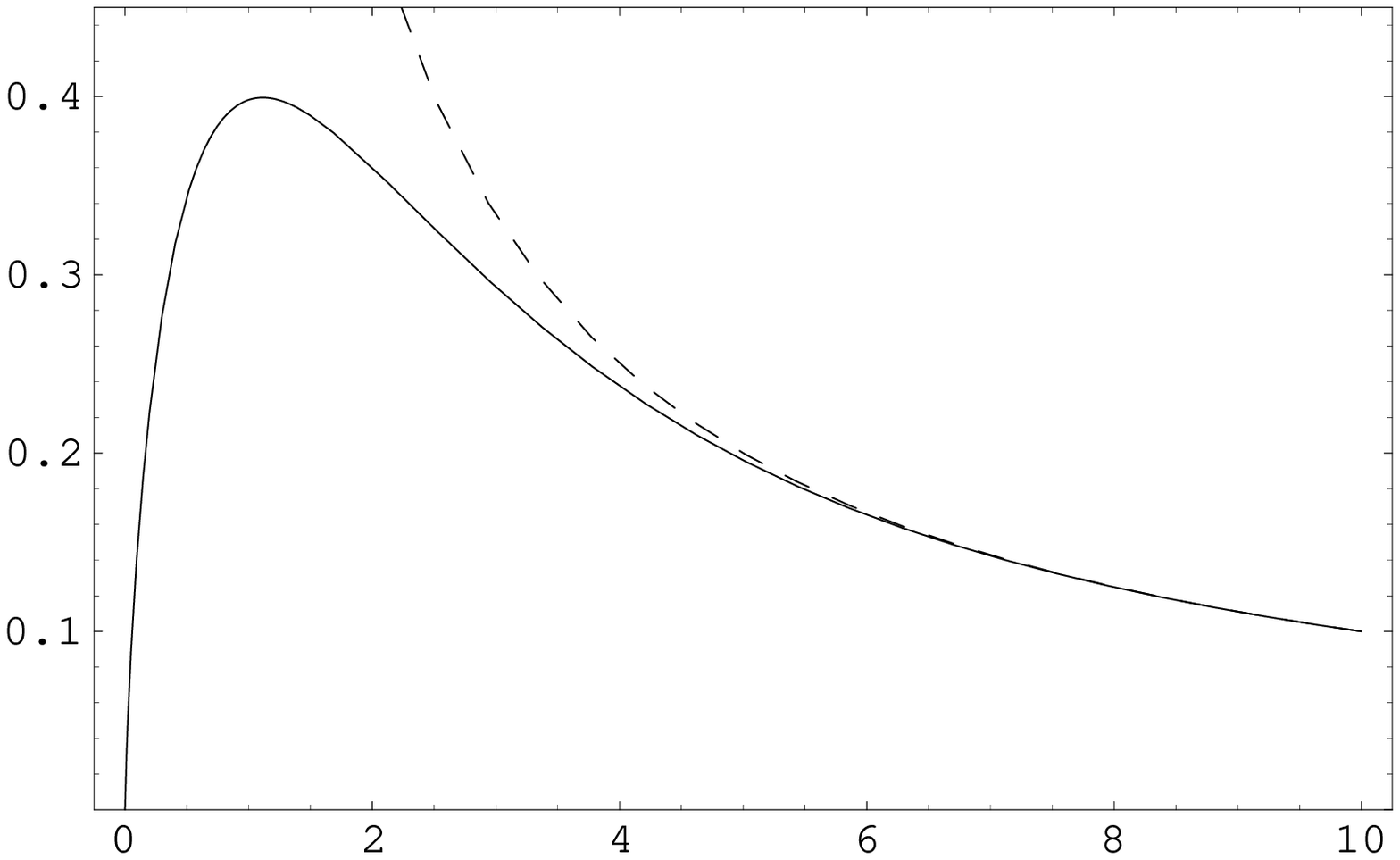,width=8cm}}
\put(4.0,0.0){$\kappa r$}
\put(-1.0,4.0){$F_r$}
\end{picture}
}
\caption{The Peach-Koehler force between screw dislocations
is given in units of $\mu \kappa b_z b'_z /[2\pi]$.
Nonlocal elasticity (solid) and classical elasticity (dashed).}
\label{fig:PKF-screw}
\end{figure}
We begin our considerations with the simple case of two parallel screw 
dislocations.
For a screw dislocation with $\xi_z=1$ we have in Cartesian coordinates
\begin{align}
&F_x=\sigma_{yz} b'_z= \frac{\mu b_z b'_z}{2\pi}\,\frac{x}{r^2}\Big\{1-\kappa r K_1(\kappa r)\Big\},\nonumber\\
&F_y=-\sigma_{xz} b'_z= \frac{\mu b_z b'_z}{2\pi}\,\frac{y}{r^2}\Big\{1-\kappa r K_1(\kappa r)\Big\},
\end{align}
and in cylindrical coordinates
\begin{align}
\label{PKF-screw-cyl}
&F_r=F_x\cos\varphi+F_y\sin\varphi
    =\frac{\mu b_z b'_z}{2\pi r}\,\Big\{1-\kappa r K_1(\kappa r)\Big\},\nonumber\\
&F_\varphi=F_y\cos\varphi-F_x\sin\varphi=0.
\end{align}
Thus, the force between two screw dislocations is also a radial force
in the nonlocal case.
The force expression (\ref{PKF-screw-cyl}) has some interesting features.
A maximum of $F_r$ can be found from Eq.~(\ref{PKF-screw-cyl}) as
\begin{align}
|F_r|_{\text{max}}\simeq 0.399\kappa\,\frac{\mu b_z b'_z}{2\pi},\quad
\text{at}\quad r\simeq 1.114\kappa^{-1}. 
\end{align}
When the nonlocal atomistic effect is neglected, $\kappa^{-1}\rightarrow 0$,
Eq.~(\ref{PKF-screw-cyl}) gives the classical result
\begin{align}
\label{PKF-screw-cyl-cl}
F^{cl}_r=\frac{\mu b_z b'_z}{2\pi r}.
\end{align}
To compare the classical force with the nonlocal one, the graphs 
from Eqs.~(\ref{PKF-screw-cyl}) and (\ref{PKF-screw-cyl-cl})
are plotted in
Fig.~\ref{fig:PKF-screw}.
It can be seen that near the dislocation line the nonlocal result 
is quite different from the classical one.
Unlike the classical expression, which diverges as $r\rightarrow 0$ and 
gives an infinite force, it is zero at $r=0$. 
For $r>6\kappa^{-1}$ the classical and the nonlocal expressions coincide.

\subsection{Parallel edge dislocations} 
We analyze now the force between two parallel edge dislocations with 
(anti)parallel Burgers vector. 
For an edge dislocation with $\xi_z=1$ with Burgers vector $b'_x$ 
we have in Cartesian coordinates
\begin{align}
\label{PKF-edge}
&F_x=\sigma_{yx} b'_x
   =\frac{\mu b_x b'_x}{2\pi(1-\nu)}\, 
\frac{x}{r^4}\bigg\{\big(x^2-y^2\big)-\frac{4}{\kappa^2r^2}\big(x^2-3y^2\big)\\
&\hspace{4cm}
-2 y^2\kappa r K_1(\kappa r)
+2\big(x^2-3y^2\big) K_2(\kappa r)\bigg\},\nonumber\\
&F_y=-\sigma_{xx} b'_x
   =\frac{\mu b_x b'_x}{2\pi(1-\nu)}\, 
\frac{y}{r^4}\bigg\{\big(y^2+3x^2\big)+\frac{4}{\kappa^2r^2}\big(y^2-3x^2\big)
\nonumber\\
&\hspace{4cm}
-2 y^2\kappa r K_1(\kappa r)
-2\big(y^2-3x^2\big) K_2(\kappa r)\bigg\}.\nonumber
\end{align}
\begin{figure}[t]\unitlength1cm
\vspace*{-1.0cm}
\centerline{
(a)
\begin{picture}(8,6)
\put(0.0,0.2){\epsfig{file=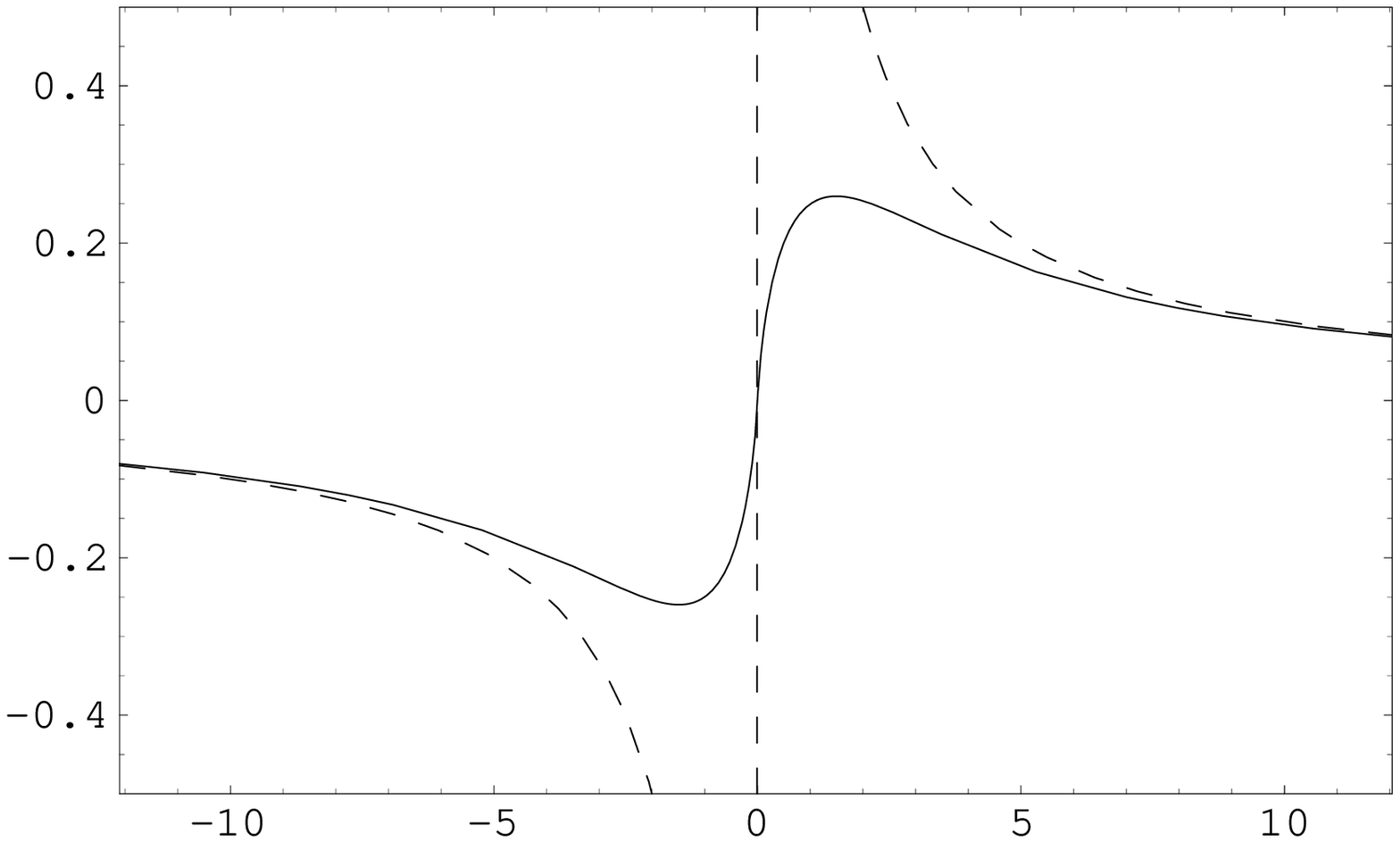,width=8cm}}
\put(4.0,0.0){$\kappa x$}
\put(-1.8,4.0){$F_{x}(x,0)$}
\end{picture}
}
\vspace*{-1.0cm}
\centerline{
(b)
\begin{picture}(8,6)
\put(0.0,0.2){\epsfig{file=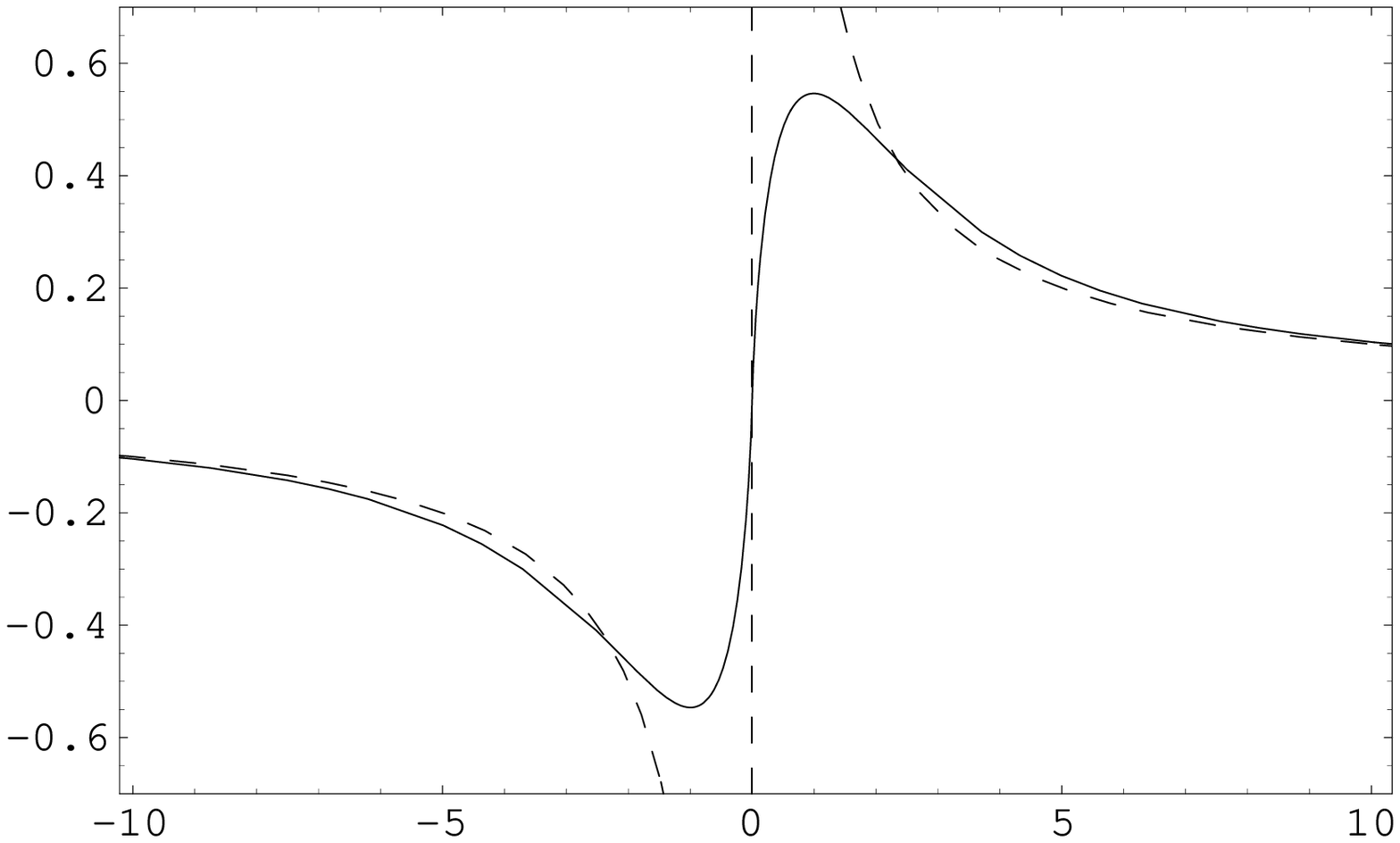,width=8cm}}
\put(4.0,0.0){$\kappa y$}
\put(-1.8,4.0){$F_{y}(0,y)$}
\end{picture}
}
\caption{The glide and climb force components near the dislocation line: 
(a) $F_{x}(x,0)$ and (b) $F_{y}(0,y)$
are given in units of $\mu b_x b'_x \kappa/[2\pi(1-\nu)]$. 
The dashed curves represent the classical force components.}
\label{fig:edge1}
\end{figure}
$F_x $ is the driving force for conservative motion (gliding)
and $F_y$ is the climb force.
The glide force has a maximum/minimum in the slip plane ($zx$-plane) of
\begin{align}
|F_x(x,0)|\simeq 0.260 \kappa\frac{\mu b_x b'_x}{2\pi(1-\nu)},
\quad\text{at}\quad |x|\simeq 1.494 \kappa^{-1}.
\end{align}
The maximum/minimum of the climb force is found as
\begin{align}
|F_y(0,y)|\simeq 0.546\kappa \frac{\mu b_x b'_x}{2\pi(1-\nu)},
\quad\text{at}\quad |y|\simeq 0.996 \kappa^{-1}.
\end{align}
It can be seen that the maximum of the climb force is greater than
the maximum of the glide force (see Fig.~\ref{fig:edge1}). 
The glide force $F_x$ is zero at $x=0$ ($\varphi=\frac{1}{2}\pi$). 
This corresponds to one
equilibrium configuration of the two edge dislocations. 
In classical elasticity the glide force is also zero at the position $x=y$ ($\varphi=\frac{1}{4}\pi$). 
But in nonlocal elasticity we obtain from Eq.~(\ref{PKF-edge}) the following 
expression for the glide force
\begin{align}
\label{PKF-edge-Fx}
F_x(\varphi=\pi/4)=\frac{\mu b_x b'_x\sqrt{2}}{4\pi(1-\nu)}\,
     \frac{1}{r}\bigg\{\frac{4}{\kappa^2 r^2}-\kappa r K_1(\kappa r)-2K_2(\kappa r)\bigg\}.
\end{align}
Its maximum is
\begin{align}
|F_x(\varphi=\pi/4)|_{\text{max}}\simeq 0.151\kappa\,\frac{\mu b_x b'_x\sqrt{2}}{4\pi(1-\nu)},\quad
\text{at}\quad r\simeq 0.788\kappa^{-1}.
\end{align}
The $F_x(\varphi=\pi/4)$ gives only a valuable contribution 
in the region $r\lessapprox 12\kappa^{-1}$.
Therefore, only for $r>12\kappa^{-1}$ the position $\varphi=\frac{1}{4}\pi$ is
an equilibrium configuration.
The climb force $F_y$ is zero at $y=0$.

From Eq.~(\ref{PKF-edge}) the force can be given 
in cylindrical coordinates as 
\begin{align}
\label{PKF-edge-cyl}
&F_r
    =\frac{\mu b_x b'_x}{2\pi(1-\nu)}\,\frac{1}{r}\bigg\{
	\Big(1-\kappa r K_1(\kappa r)\Big)\nonumber\\
&\hspace{3cm}     
-\cos 2\varphi\bigg(\frac{4}{\kappa^2 r^2}-\kappa r K_1(\kappa r)-2K_2(\kappa r)\bigg)\bigg\},\nonumber\\
&F_\varphi
=\frac{\mu b_x b'_x}{2\pi(1-\nu)}\,\frac{\sin 2\varphi}{r}
\bigg\{1-\frac{4}{\kappa^2 r^2}+2K_2(\kappa r)\bigg\}.
\end{align}
The force between edge dislocations is not a central force because
a tangential component $F_\varphi$ exists. 
Both the components $F_r$ and $F_\varphi$ depend on $r$ and $\varphi$.
The dependence of $\varphi$ in $F_r$ is a new feature of the
nonlocal result~(\ref{PKF-edge-cyl}) not present in the classical elasticity. 
Unlike the ``classical'' case,
the force $F_r$ is only zero at $r=0$ and $F_\varphi$ is zero at $r=0$ and
$\varphi=0,\,\frac{1}{2}\pi,\,\pi,\, \frac{3}{2}\pi,\, 2\pi$.
The force $F_r$ in~(\ref{PKF-edge-cyl}) has an interesting dependence of 
the angle $\varphi$. In detail, we obtain 
\begin{alignat}{2}
&\text{(i)}\
\varphi=0,\,\pi:\quad
&&F_r=\frac{\mu b_x b'_x}{2\pi(1-\nu)}\,\frac{1}{r}
\bigg\{1-\frac{4}{\kappa^2 r^2}+2K_2(\kappa r)\bigg\},\\
& && |F_r|_{\text{max}}\simeq 0.260\,\frac{\mu b_x b'_x\kappa}{2\pi(1-\nu)},\quad
\text{at}\quad r\simeq 1.494\kappa^{-1},\nonumber
\end{alignat}
\begin{alignat}{2}
&\text{(ii)}\
\varphi=\frac{\pi}{4},\,\frac{3\pi}{4},\, \frac{5\pi}{4},\, \frac{7\pi}{4}:\ 
&&F_r=\frac{\mu b_x b'_x}{2\pi(1-\nu)}\,\frac{1}{r}
\big\{1-\kappa rK_1(\kappa r)\big\},\\
& && |F_r|_{\text{max}}\simeq 0.399\,\frac{\mu b_x b'_x\kappa}{2\pi(1-\nu)},\ \ 
\text{at}\ \  r\simeq 1.114\kappa^{-1},\nonumber
\end{alignat}
\begin{alignat}{2}
&\text{(iii)}\
\varphi=\frac{\pi}{2}, \frac{3\pi}{2}:\ 
&&F_r=\frac{\mu b_x b'_x}{2\pi(1-\nu)}\frac{1}{r}
\bigg\{1+\frac{4}{\kappa^2 r^2}-2\kappa rK_1(\kappa r)-2K_2(\kappa r)\bigg\},\nonumber\\
& && |F_r|_{\text{max}}\simeq 0.547\,\frac{\mu b_x b'_x\kappa}{2\pi(1-\nu)},\ \ 
\text{at}\ \ r\simeq 0.996\kappa^{-1}.
\end{alignat}
\begin{figure}[t]\unitlength1cm
\vspace*{-1.0cm}
\centerline{
(a)
\begin{picture}(8,6)
\put(0.0,0.2){\epsfig{file=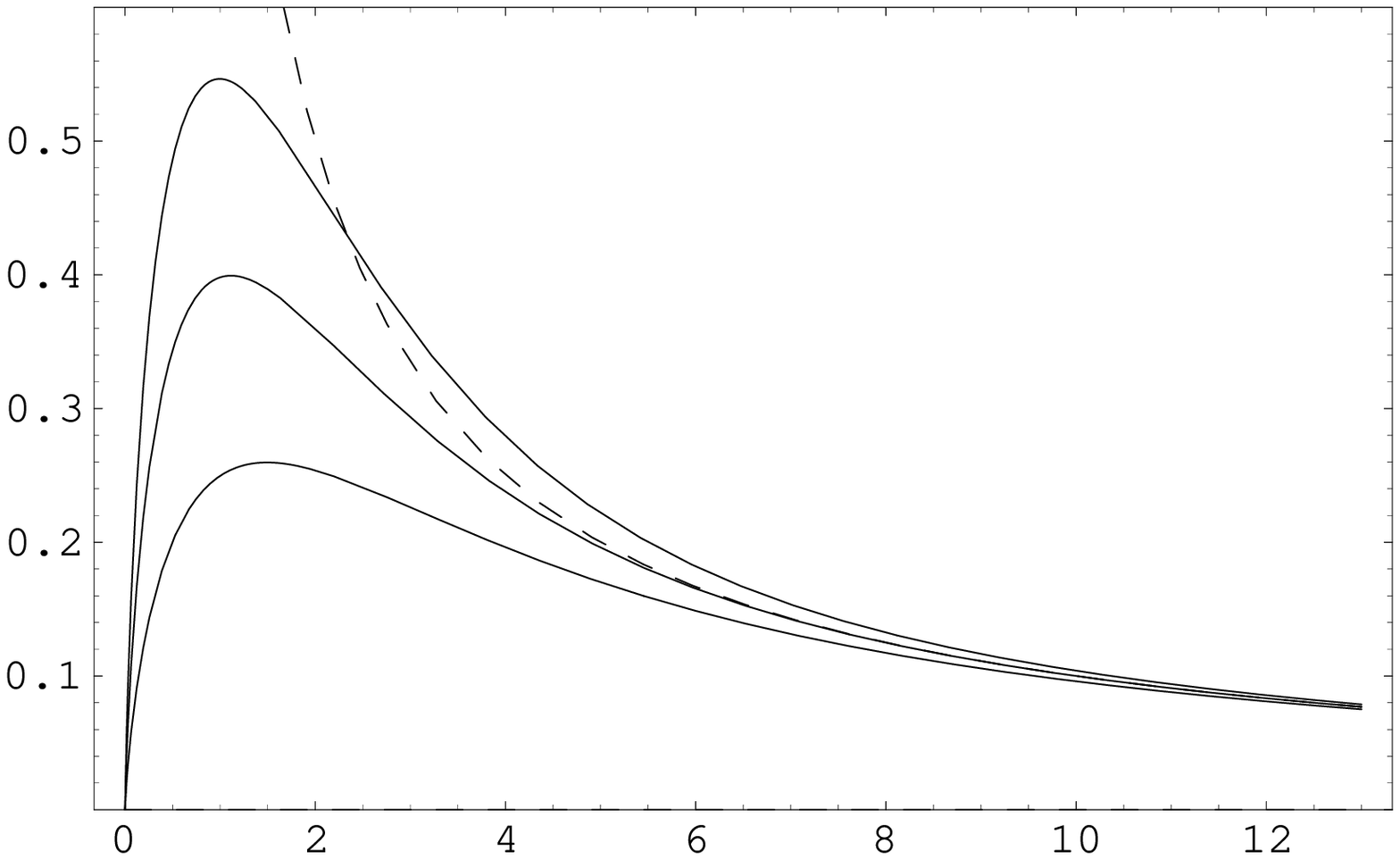,width=8cm}}
\put(4.0,0.0){$\kappa r$}
\put(-1.5,4.0){$F_{r}$}
\put(1.2,1.9){(i)}
\put(1.1,2.9){(ii)}
\put(1.05,3.9){(iii)}
\end{picture}
}
\vspace*{-1.0cm}
\centerline{
(b)
\begin{picture}(8,6)
\put(0.0,0.2){\epsfig{file=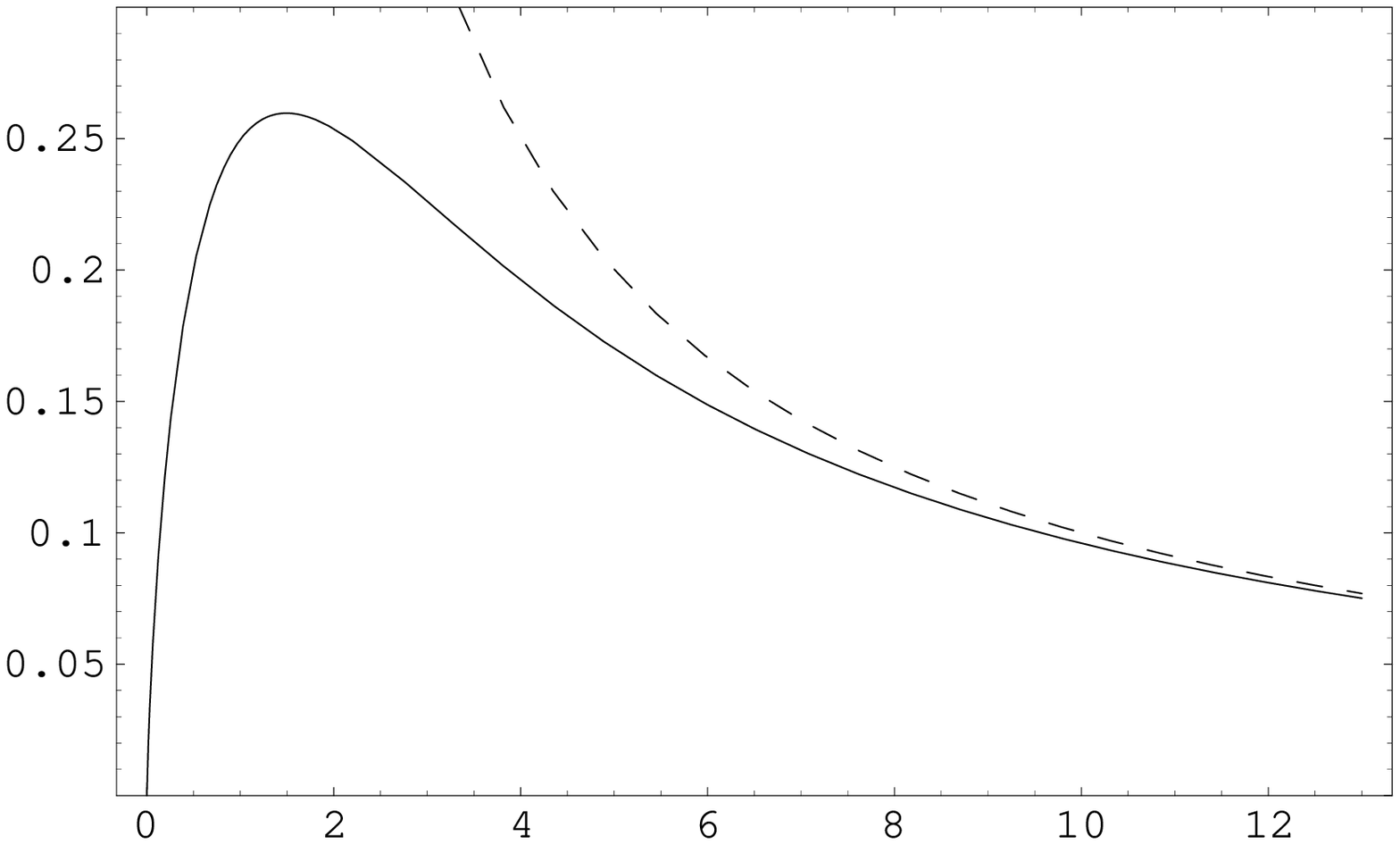,width=8cm}}
\put(4.0,0.0){$\kappa r$}
\put(-1.5,4.0){$F_{\varphi}$}
\end{picture}
}
\caption{The force near the dislocation line: 
(a) $F_{r}$ and (b) $F_{\varphi}$ with $\varphi=\frac{1}{4}\pi,\, \frac{5}{4}\pi$
are given in units of $\mu b_x b'_x \kappa/[2\pi(1-\nu)]$. 
The dashed curves represent the classical force components.}
\label{fig:edge2}
\end{figure}
On the other hand, $F_\varphi$ has a maximum at $\varphi=\frac{1}{4}\pi, \frac{5}{4}\pi$ and
a minimum at $\varphi=\frac{3}{4}\pi, \frac{7}{4}\pi$.
They are 
\begin{align}
|F_\varphi|\simeq 0.260\kappa \frac{\mu b_x b'_x}{2\pi(1-\nu)},
\quad\text{at}\quad r\simeq 1.494 \kappa^{-1}.
\end{align}
The classical result for the force reads in cylindrical coordinates
\begin{align}
\label{PKF-edge-cyl-cl}
&F^{cl}_r=\frac{\mu b_x b'_x}{2\pi(1-\nu)}\,\frac{1}{r},\quad
F^{cl}_\varphi
=\frac{\mu b_x b'_x}{2\pi(1-\nu)}\,\frac{\sin 2\varphi}{r}.
\end{align}
To compare the nonlocal result with the classical one, diagrams of
Eqs.~(\ref{PKF-edge-cyl}) and (\ref{PKF-edge-cyl-cl}) are drawn in 
Fig.~\ref{fig:edge2}.
The force calculated in nonlocal elasticity
is different from the classical one
near the dislocation line at $r=0$. It is finite and has
no singularity in contrast to the classical result.
For $r>12\kappa^{-1}$ the classical and the nonlocal expressions coincide.

\section{Conclusion}
The nonlocal theory of elasticity has been used to calculate the Peach-Koehler force
due to screw and edge dislocations in an infinitely extended body.
The Peach-Koehler force calculated in classical elasticity is infinite near 
the dislocation core. The reason is that the classical elasticity is invalid in dealing with problems of
micro-mechanics. The nonlocal elasticity gives expressions for the
Peach-Koehler force which are physically more reasonable. 
The forces due to dislocations have no singularity. They are zero at $r=0$
and have maxima or minima in the dislocation core. 
The finite values of the force due to dislocations may be used to analyze
the interaction of dislocations from the micro-mechanical point of view.

\begin{acknowledgments}
This work was supported by the European Network TMR 98-0229.
\end{acknowledgments}


\end{document}